\documentclass[aps,pra,reprint]{revtex4-1}

\usepackage{amsmath}
\usepackage{graphicx}
\usepackage{dcolumn}
\usepackage{verbatim}

\newcolumntype{d}{D{.}{.}{-1}}

\begin{document}

\title{The Variationally Fitted  Electron-Electron Potential}
\author{Brett I. Dunlap}
\email{dunlap@nrl.navy.mil}
\author{Mark C Palenik}
\thanks{NRC post-doctoral fellow}
\address{Code 6189, Chemistry Division, Naval Research
	Laboratory, Washington, DC 20375, United States}

\begin{abstract}
	
Perhaps the simplest first-principles approach to electronic structure is to fit the charge distribution of each orbital pair and use those fits wherever they appear in the entire electron-electron (EE) interaction energy. The charge distributions in quantum chemistry are typically represented as a sums over products of Gaussian orbital basis functions.  If fitted, they are also represented as a sum over single-center Gaussian fitting basis functions.  With two representations of the charge distributions, the proper definition of energy is ambiguous.  To remedy this, we require that the variation of the energy with respect to a product of orbitals generates a fitted potential.  This makes the quantum-mechanical energy robust, i.e. corrected to first order for the error made using an incomplete  fitting basis.  The coupled orbital and fitting equations are then the result of making the energy stationary with respect to two independent sets of variables.  We define the potentials and unique energies for methods based on the Hartree Fock model and variationally fit the full EE interaction in DFT.  We compare implementations of variational fitting in DFT at six different levels for three different functionals.  Our calculations are performed on transition metal atoms, for which first-order Coulomb errors, due to an incomplete fitting basis sets, are significant.  Variational first-order exchange and correlation errors have similar magnitude in all cases.  Robust energy differences are much smaller, particularly in the local density approximation.

\end{abstract}

\maketitle

\section{Introduction}

The computationally challenging part of electronic-structure calculations is obtaining the self-consistent field (SCF) that makes the electron-electron (EE)  interaction energy  stationary.  This drives interest in density functional theory (DFT).  In DFT, the electronic energy and density are variationally linked according to the Hohenberg-Kohn (HK) theorems \cite{HKTheorems}, which prove that the ground-state density minimizes the ground-state energy.  In Kohn-Sham (KS) DFT, this is realized through the KS potential, which results from minimizing the EE interaction energy with respect to the density \cite{Kohn1965}, because the kinetic energy is taken to be that of non interacting electrons.

The computation of the DFT energy is only simplified relative to the equivalent Gaussian-orbital-based Hartree-Fock (HF) calculations if the KS potential is fitted \cite{SambeFelton1975}, yielding methods that scale as the number of electrons to the third power.  When variational fitting is applied to the Coulomb potential, it becomes a function of the fitted electron density \cite{Dunlap1979,Dunlap1979-2}.   Therefore variational fitting is often, but not universally, called density fitting, but it is the fitted potential that simplifies the computation of the molecular orbitals and energy.  The fitted density is generated collaterally.   The same fitting equations can be used to fit neutral potentials to a sum of short-ranged Gaussians \cite{Mintmire1982}.  In that case, too, it is the special properties of the fitted potential and not those of the fitted density (other than it required neutrality) that are the reasons for fitting.

The purpose of this work is to simultaneously fit the charge distributions and corresponding potentials for all of electronic structure theory, which is perhaps the simplest possible approach to quantum chemistry.   A unique energy can be defined in terms of the fitted potential and fitted and exact charge densities upon which the full calculus of variations (CoV) can be applied in both the orbital and fitting spaces.

The number of ways to fit electronic charge distributions using fitting basis sets is likely on the order of the number of scientists who have practiced quantum chemistry.  Certainly the practice dates back to the beginning of quantum chemistry \cite{Shavitt1959}.  With so many practitioners, the differences can be quite small, but not insignificant as one desires to treat ever larger systems.  

In our view, the variational relationship between the energy and the wavefunction is central to quantum mechanics.  In KS DFT the variational relationship is between the energy and the KS orbitals.  In HK DFT, it is between the energy and the density.   In all cases, when fitting introduced, it seems computationally advantageous to define the energy is such a way that its variational properties are preserved.  Doing so both simplifies the EE potential (and thus the calculation) {\it and} preserves its variational relationship with the energy.  We call approximations that preserve the variational principle variational fitting \cite{Dunlap2010Mol}.

Under very special circumstances, the difference between EE energy evaluated using {\it only}  fitted charge distributions and the variationally fitted energy that involves both fitted and exact charge distributions is numerically zero \cite{Whitten1973}.   Even for that special case, if the fit is constrained to generate reasonable long-range interaction energies, then  the variational fitted energy is different.  Motivating our work is the fact the difference in energy using constrained and unconstrained Coulomb fits with reasonable fitting basis sets is very large in calculations on heavy atoms \cite{Dunlap1977}, but variational fitting through the CoV largely zeros that difference \cite{Dunlap1979}.  It also largely zeros the energy difference if the entire EE interaction energy is fitted, as we will show.

Uses of the Coulomb fitting equation (Eq.~(\ref{Eq5}) below) have been given various names by the developers of commercial quantum-chemistry software \cite{Feyereisen1993,Vahtras1993,Aquilante2008,Yang2007}.  Overall, perhaps the number of new names is approaching the number of workers in the field \cite{Boman2008}.  We are aware of no discussion of why this work should not be called variational fitting.  Perhaps the different names are due to the fact that only Coulomb fitting basis sets have been optimized using the CoV \cite{Godbout1992}.  Fitting basis sets for the other methods are not variational, like the non-variational methods used to develop early Gaussian orbital basis sets \cite{Huzinaga1965,Huzinaga1986}.  Those early orbital basis sets have been replaced by variational Pople basis sets \cite{Ditchfield1971}.  The new name Resolution of the Identity (RI) has been subdivided.  In RI-J one is fitting the density \cite{Eichkorn1995,Eichkorn1995Err}, in RI-K one is fitting all pairs of occupied orbitals that are different \cite{Weigend2008}, and in RI-MP2 one is fitting all pairs of occupied and virtual orbitals \cite{Weigend1998,Hellweg2007}.  All these fits are of charge distributions.  Those charge distributions are fitted in the same way we simultaneously fit the density and Coulomb potential.  They are all unconstrained fits, which creates ambiguity in the definition of the energy \cite{Dunlap2000}.

The EE interaction in the Fock matrix is the expectation value of the HF potential in the orbital basis.  The Gaussian products of orbital basis function pairs, which are generated by minimizing the HF energy, cannot be separated into orbital-same-orbital, orbital-other-orbital, and occupied-virtual basis-function pairs corresponding to the three named fits.  The time-consuming step in RI-K is transforming the fitted two-electron integrals into exchange matrix elements.    Thus RI-HF scales as the fourth power of the number of electrons.  This poor scaling is a part of modern density functional theory (DFT) \cite{Manzer2015}.   Two-electron integral transformations would not be required if exact exchange were treated in perturbation theory \cite{Palenik2015}.

Another alternative to the integral transformation step of RI-K would be to treat the orbital pairs, rather than orbitals, as the functions that minimize the energy.  The same construction that was used for KS DFT can be applied in HF to define the corresponding potential.  We take the kinetic energy to be that of noninteracting particles and vary the HF EE energy with respect to the orbital-pair charge distributions.  The entire electronic structure problem is reformulated as a set of non-interacting electrons to simultaneously determine the kinetic and EE energies.  The energy is minimized by variation with respect to the orbital pairs evaluated at a single point.  Fitting the resultant potentials of a single variable reduces the scaling of the calculation of these matrix elements of the potential to below the third power of the number of electrons if localized fitting basis sets are used.

The matrix elements of these potentials go directly, without an integral transformation step, into the Fock matrix that determines the orbitals.  In DFT one is not interested in solving for the HF energy, but exact exchange is a reasonable thing to add to DFT.   If the electronic structure methods are mixed and both the KS potential and the HF matrix elements are fitted, then the fits interact.   This interaction has not been studied.  Using the CoV, we determine the equation that couples the fits.  The same equation couples the fits to the Coulomb and exchange and correlation (XC) parts of the KS potential.  That coupling is numerically studied in this work.  The variational principle largely cancels the two errors in the potentials even though the Coulomb energy (CE) is an order of magnitude larger than the XC energy.   Thus, we expect that using the fitted density from RI-HF calculations to generate the KS potential in modern DFT, would create a much simpler and equally accurate calculation.  Of course, such calculations would still be slower than the non-modern DFT calculations that only require a KS potential.

The next section views the EE interaction energy as the result of a set of charge distributions created by pairs of orbitals interacting with each other.  It defines the potential as the sum of the variations of the EE energy with respect to each orbital pair.   In the third section, we apply that formalism to HF methods to define the relevant potentials and, through them, the unique fitted HF energies that can be fully optimized using the CoV.  In the fourth section we use the KS potential generated by the fitted density to obtain the variational fitted DFT energy.   In the fifth section we demonstrate the exquisite power of the CoV by using it alone to fit the KS potential and density in perhaps the most efficient possible Gaussian-basis-set  generalized-gradient-approximation (GGA) DFT calculations on heavy atoms.   By construction, these new methods have no first-order error due to fitting.

\section{Simultaneously fitting both orbital charge distributions and their potentials}
Varying the EE energy with respect to the product of two orbitals $u^*_i(\mathbf{r})u_j(\mathbf{r})$ generates a potential of a single variable, $V_{ij}(\mathbf{r})$.  Collectively they define the EE potential,
\begin{equation}
\begin{split}
V_{ee}=&\sum_{ij} {| u_i\rangle} \int V_{ij}(\mathbf{r})
\rho_{ij}(\mathbf{r})d\mathbf{r} {\langle u_j |} 
\\=&
\sum_{ij} {| u_i\rangle} \int\frac {\delta E_{ee}[\rho]} {\delta\rho_{ij}(\mathbf{r})} 
\rho_{ij}(\mathbf{r})d\mathbf{r} {\langle u_j |} 
\end{split}
\label{Eq1}
\end{equation}
and the $V_{ij}(\mathbf{r})$ are functions of the charge distributions of the various orbital pairs.

The expectation value of the potential in Eq.~(\ref{Eq1}) is not equal to the EE energy.  Because the energy is non-linear it requires a double-counting (DC) correction.  The name DC correction comes from the CE, which differs from the expectation value of the Coulomb potential by a factor of one-half, although this is not the case in general.  When fitting is introduced, there is some ambiguity in the definition of the energy, because it is not clear where, other than in the potentials, fitted densities should be used.

This ambiguity can be removed by requiring that the variational principle we have outlined relating the energy and potentials to hold.  When this is the case, both the potential and the DC correction must be expressed in terms of fitted density.  Using a fitted DC term makes the energy robust because it is corrects the energy to first order for errors introduced by using two different representations of the density \cite{Dunlap2000}.  In general, the robust energy can be written as
\begin{equation}
E[\rho,\bar{\rho}]=\int V[\bar{\rho}(\mathbf{r})]\rho(\mathbf{r})d\mathbf{r}+E_{DC}[\bar\rho]
\label{Eq2}
\end{equation}
Applying the CoV to the robust energy must yield the potential once again.  In order for that to happen the chain rule requires the identity,
\begin{equation}
0=\frac{\delta
V[\bar\rho(\mathbf{r})]}{\delta\bar\rho(\mathbf{r})}\rho(\mathbf{r})+\frac{\delta
E_{DC}[\bar\rho]}{\delta\bar\rho(\mathbf{r})}
\label{Eq3}
\end{equation}	
This equation can be used independently of whether or not the potential is an analytic function of the density.  It will be used in numerical DFT calculations later.

Applying these principles to the Coulomb energy, it is expressed as the expectation value of the Coulomb potential plus a double counting term,
\begin{equation}
\begin{split}
E_{CE}&=\int \bar{V}_{CE}(\mathbf{r})\rho(\mathbf{r}) d\mathbf{r}+ \bar{E}_{DC}\\
&=\langle\rho|\bar\rho\rangle-\textstyle{1\over 2}\langle\bar\rho|\bar\rho\rangle
\end{split}
\label{Eq4}
\end{equation}
where $\bar{V}_{CE}$ indicates that the Coulomb potential is a function of the fitted density, and the vertical line indicates the Coulomb interaction between the pair of charge distributions on either side. Applying Eq.~(\ref{Eq3}) within these integrals, we get
\begin{equation}
\langle\rho-\bar\rho|\delta\bar\rho\rangle=0
\label{Eq5}
\end{equation}
The CoV applied to this equation determines the linear fitting
coefficients and non-linear basis-function exponents that make the robust energy
stationary \cite{Dunlap1979,Dunlap1979-2}.  Complete variational freedom makes the two
representations of the density equal.

\section{Robust Fitted Hartree-Fock Energies and Their Variations}
The EE energy in HF is a function of its two electron integrals.  Defining an orbital charge distribution, $\rho_{ij}(\mathbf{r})=u^*_i(\mathbf{r})u_j(\mathbf{r})$, we can write this energy, in the notation of Eq.~(\ref{Eq4}) as
\begin{equation}
E_{ee} = \textstyle{1\over2}\sum_{i,j}\langle\rho_{ii}|\rho_{jj}\rangle-\langle\rho_{ji}|\rho_{ij}\rangle
\end{equation}
where the summation is over occupied orbitals.  From this, we can define the potentials, $V_{kl}(\mathbf{r})$ that act on each charge distribution
\begin{equation}
V_{kl}(\mathbf{r})\equiv \frac{\delta E_{ee}}{\delta\rho_{kl}(\mathbf{r})} = {\delta_{kl}\sum_iU_{ii}(\mathbf{r})-U_{lk}(\mathbf{r})}
\end{equation}
where $U_{lk}(\mathbf{r})$ is the Coulomb potential due to charge distribution $\rho_{lk}(\mathbf{r})$.  The EE energy is given by the sum of the matrix elements of the orbital potentials with the corresponding orbitals plus a DC correction.  In the robust EE energy, the potential and the DC correction are functions of fitted orbital charge distributions, $\bar\rho_{ij}$, yielding
\begin{equation}
\begin{split}
&E_{ee} = \sum_{i,j}\int\bar{V}_{ij}(\mathbf{r})\rho_{ij}(\mathbf{r})d\mathbf{r} + \bar{E}_{DC}
\\&=\sum_{i,j}\left[\langle\bar{\rho}_{ii}|\rho_{jj}\rangle-\langle\bar{\rho}_{ji}|\rho_{ij}\rangle- \textstyle{1\over2}\left(\langle\bar{\rho}_{ii}|\bar\rho_{jj}\rangle-\langle\bar{\rho}_{ji}|\bar\rho_{ij}\rangle\right)\right]
\end{split}
\label{Eq8}
\end{equation} 
The CoV gives the fitting equation
\begin{equation}
0=\sum_j\langle\rho_{jj}-\bar{\rho}_{jj}|\delta\bar\rho_{ik}\rangle\delta_{ik}-\langle\rho_{ki}-\bar{\rho}_{ki}|\delta\bar\rho_{ik}\rangle
\label{Eq9}
\end{equation}
The presence of $\delta_{ik}$ means that first term only contributes when $i=k$.  Those diagonal terms do not include self-interaction, and this differs from RI-J unless the same basis is used in all fits.  For exchange, $i \neq k$ and using unconstrained fits and a global fitting basis, the RI-K fits and the fits of Eq.~(\ref{Eq5}) for all occupied orbital-other-orbital charge distributions would satisfy this equation. Of course for large systems, one needs constrained fits and one would like to use a fitting basis set localized about the centroid of $\rho_{ik}(\mathbf{r})$, which becomes possible by using a localized basis for $\bar\rho_{ik}(\mathbf{r})$ in Eq.~(\ref{Eq9}).  The off-diagonal fits of this equation can be treated in parallel with added computation cost similar to that associated with unfitted direct SCF methods \cite{Almlof1982}.

In perturbation theory, the second-order HF energy \cite{Moller1934} can be written and reexpressed for variation \cite{Szabo1989modern} as
\begin{equation}
\begin{split}
E^{(2)}&=-\sum_{a<b,i<j}\frac{|\langle\rho_{ai}|\rho_{bj}\rangle-\langle\rho_{aj}|\rho_{bi}\rangle|^2}{\epsilon_a+\epsilon_b-\epsilon_i-\epsilon_j}
\\&=-\frac{1}{2}\sum_{a,i,b,j}\langle \rho_{ai}|\rho_{bj}\rangle\frac{\langle \rho_{ia}|\rho_{jb}\rangle - \langle \rho_{ib}|\rho_{ja}\rangle}{\epsilon_a+\epsilon_b-\epsilon_i-\epsilon_j}
\end{split}
\label{Eq10}
\end{equation}
where $i,j$ denote occupied orbitals and $a,b$ denote virtual HF orbitals.  We define potentials for this energy through the CoV,
by varying the energy with respect to the occupied-virtual orbital-product charge distributions.
\begin{equation}
\begin{split}
V_{ck}(\mathbf{r})&\equiv\frac{\delta E^{(2)}}{\delta
\rho_{ck}(\mathbf{r})}\\
&=-\sum_{a,i}{U_{ai}(\mathbf{r})}\frac{\langle \rho_{kc}|\rho_{ia}\rangle - \langle \rho_{ka}|\rho_{ic}\rangle}{\epsilon_c+\epsilon_a-\epsilon_k-\epsilon_i}
\end{split}
\end{equation}
and $V_{kc}(\mathbf{r})$, the variation of $E$ with respect to $\rho_{kc}(\mathbf{r})$, is its complex conjugate.

The robust energy can then be written as the expectation value of the these potentials with the appropriate orbital products, plus the DC correction, where both the potentials and DC correction are in terms of the fitted orbital charge distributions
\begin{equation}
\begin{split}
E^{(2)} =& 2\mathrm{Re}\sum_{c,k}\int\bar{V}_{ck}(\mathbf{r})\rho_{ck}(\mathbf{r}) d\mathbf{r} + \bar{E}_{DC}\\
=& 2\mathrm{Re}\sum_{c,k}\int\bar{V}_{ck}(\mathbf{r})\rho_{ck}(\mathbf{r}) d\mathbf{r} 
\\&+\frac{3}{2}\sum_{a,i,b,j}\langle \bar\rho_{ai}|\bar\rho_{bj}\rangle\frac{\langle \bar\rho_{ia}|\bar\rho_{jb}\rangle - \langle \bar\rho_{ib}|\bar\rho_{ja}\rangle}{\epsilon_a+\epsilon_b-\epsilon_i-\epsilon_j}
\end{split}
\label{Eq12}
\end{equation}
Applying the CoV to this equation gives the solution,
\begin{equation}
0=\int (\rho_{ck}-\bar\rho_{ck})\delta\bar{V}_{ck}d\mathbf{r}
\label{Eq13}
\end{equation}
for each product of orbitals appearing in Eq.~(\ref{Eq12}).   With unconstrained fits and a global fitting basis the RI-MP2 fits and the fits of Eq.~(\ref{Eq5}) for all occupied-virtual charge distributions would satisfy this equation.   Of course, for large systems, one needs constrained fits and one would like to use a fitting basis set localized about the centroid of $\rho_{ck}(\mathbf{r})$.  In either case, only the robust energy has no first-order error.

\section{Exchange and Correlation in General}

The general, DFT exchange and correlation (XC) energy density is a functional of the
spin density, gradients and Laplacians of those densities, {\it etc}., up Perdew''s
ladder of functionals \cite{Perdew2005}.   As the functionals are quite varied, we express the DC
 in terms of the XC energy density, $\varepsilon_{XC}[\rho]$, and potential,
$V_{XC}[\rho]$ according to Eq.~(\ref{Eq2}),
\begin{equation}
\begin{split}
E_{XC} &= \int V_{XC}[\bar{\rho}(\mathbf{r})]\rho(\mathbf{r})d\mathbf{r} + \bar{E}_{DC}[\bar{\rho}]
\\&= \int\varepsilon_{XC}[\bar\rho(\mathbf{r})]+V_{XC}[\bar\rho(\mathbf{r})]
[\rho(\mathbf{r})d\mathbf{r}-\bar\rho(\mathbf{r})]d\mathbf{r}
\label{Eq14}
\end{split}
\end{equation}
where the XC energy-density and potential are only determined using the fitted
density. 

In general, there is a functional derivative or potential for every
independent variable of every rung.  In the second rung, the GGA, the
independent variables are both spin densities and $\sigma_1=|\nabla\rho_{\uparrow}(\mathbf{r})|^2$, $\sigma_2=\nabla\rho_{\uparrow}(\mathbf{r})\cdot\nabla\rho_{\downarrow}(\mathbf{r})$, and $\sigma_3=|\nabla\rho_{\downarrow}(\mathbf{r})|^2$, and the potentials in
Eq.~(\ref{Eq14}) are operators that can be combined.  For spin up, the XC
potential operator is
\begin{equation}
\hat{V}_{XC\uparrow}=V_\uparrow(\mathbf{r})+\left[2V_{\sigma_1}(\mathbf{r})\nabla\bar\rho_\uparrow(\mathbf{r})+V_{\sigma_2
}(\mathbf{r})\nabla\rho_\downarrow(\mathbf{r})\right]\cdot\nabla
\label{Eq20}
\end{equation}
and a similar expression results for spin down.   Applying the CoV to the fits of
Eqs.~(\ref{Eq5}) and (\ref{Eq14}),
\begin{equation}
\begin{split}
0&=\int d\mathbf{r'}\delta\bar{\rho}_j(\mathbf{r'})\left\{\bar{V}_j(\mathbf{r'})+d\mathbf{r}\sum_i \right.\\
&\left.\frac{\delta
	\bar{V}_i(\mathbf{r})}{\delta\bar\rho_j(\mathbf{r'})}\left[\rho_i(\mathbf{r})-\bar\rho_i(\mathbf{r})\right]-\bar{V}_i(\mathbf{r})\frac{\delta\bar\rho_i(\mathbf{r})}{\delta\bar\rho_j(\mathbf{r'})}\right\}
\end{split}
\label{Eq16}
\end{equation}
where subscripts $i$,$j$ indicate independent variables from each rung on
Perdew''s ladder of functionals and $V_i$ is the total correspond potential.  The first term and the last term within curly brackets cancel for the Coulomb potential and we are left with Eq.~(\ref{Eq5}) expressed in terms of the density and potential, rather than just in terms of the density.   Similarly, those two terms cancel for local XC functionals and only variations of the potential remain.  The robust HF EE energy, Eq.~(\ref{Eq8}) or equivalently its exchange component can be mixed into the energy.   Then the CoV would give the the same equation, but the sum over $i$ would extend over all orbital-other-orbital charge distributions.  Climbing the ladder or adding the robust HF second-order energy, Eq.~(\ref{Eq12}), to the energy would extend the sum over $i$ to all occupied-virtual charge distributions.  Note that in all cases this equation couples fits.   Variational fitting does not allow individual fits of charge distributionss because each fit contributes to the total potential, which modifies the other orbital charge distributions.   Numerical and incomplete-basis-set solutions to this equation are analyzed in the next section.  Solutions of this equation are obtained by simultaneous Newton-Raphson (NR) and SCF iterations.

\section{Heavy-Atom Calculations}
All practical Gaussian fitting basis sets are incomplete, introducing an error that is
significant for transition metal atoms \cite{Dunlap1977},  which are the focus of a few test
calculations.  No one has experience with variational fitting that includes XC
or with the robust XC energy of Eq.~(\ref{Eq14}), and it is the purpose of this work
to gain a little.   Similar calculations do exist.  In the model-potential method of ParaGauss,\cite{Birkenheuer2005}  the normalized,
Coulomb-fitted density is used in the XC energy density.   The chain rule, with
a Lagrange multiplier to enforce normalization, is used to obtain the entire KS
potential as a function of the fitted density.   These equations differ from Eq.~(\ref{Eq16}) in
that the only fitting is done via Eq.~(\ref{Eq5}).   Such calculations were reproduced and are called constrained variational CE in the following.   Standard fitting basis sets
were not found to be sufficiently accurate in the ParaGauss work.  Similarly, the auxiliary
  density functional theory (ADFT) of deMon2k, uses the same general approach, but
  the density is not normalized, which seems to work better.    Such calculations were reproduced and are called unconstrained variational CE in the following. The
  density fits of both the model potential of ParaGauss and ADFT make the CE and
  not the total EE energy stationary.  The EE must be modified to make
  the total DFT energy robust in both cases.  Some properties of that correction and its variational treatment are studied in this section.

\begin{table*}[t]
	\caption{Robust, relative atomic Zn energies.  For each functional, VWM, PBE,
		and BLYP, the Coulomb, CE, total electron-electron, EE, and total energies, in
		Hartree, are given relative to SCF energies computed with variationally fitted
		Coulomb energy and the XC energy of the true density.}
	\centering
	\begin{ruledtabular}
		\begin{tabular}{l c| d d d| d d d| d d d}
			&&\multicolumn{3}{c|}{VWN}&\multicolumn{3}{c|}{PBE}&\multicolumn{3}{c}{BLYP}\\
			\cline{3-11}
			Fitting Method&Constraint&\multicolumn{1}{c}{CE}& \multicolumn{1}{c}{EE}&\multicolumn{1}{c|}{Total}&\multicolumn{1}{c}{CE}&\multicolumn{1}{c}{EE}&\multicolumn{1}{c|}{Total}&\multicolumn{1}{c}{CE}&\multicolumn{1}{c}{EE}&\multicolumn{1}{c}{Total}\\  
			\hline
			Numeric& Yes&-0.002&-0.002&0.000&-0.002&-0.002&0.000&-0.002&-0.002&0.000\\
			\hline
			Variational&Yes&0.053&0.056&-0.003&0.079&0.126&0.047&-0.222&0.348&0.084\\
			Densities&No&0.053&0.057&-0.003&0.070&0.072&0.004&-0.197&0.273&0.125\\
			\hline
			Variational& Yes&-0.125&-0.112&0.004&-0.167&-0.155&0.003&-0.102&-0.097&-0.001\\
			Coulomb&No&-0.056&-0.046&0.004&-0.108&-0.098&0.003&-0.047&-0.046&-0.002\\
		\end{tabular}
	\end{ruledtabular}
	\label{Table1}
\end{table*}

\begin{table*}[t]
	\caption{Robust, relative atomic Zn energies.  For each functional, VWM, PBE,
		and BLYP, the Coulomb, CE, total electron-electron, EE, and total energies, in
		Hartree, are given relative to SCF energies computed with variationally fitted
		Coulomb energy and the XC energy of the true density.}
	\centering
	\begin{ruledtabular}
		\begin{tabular}{l c| d d d| d d d| d d d}
			&&\multicolumn{3}{c|}{VWN}&\multicolumn{3}{c|}{PBE}&\multicolumn{3}{c}{BLYP}\\
			\cline{3-11}
			Fitting Method&Constraint&\multicolumn{1}{c}{CE}& \multicolumn{1}{c}{EE}&\multicolumn{1}{c|}{Total}&\multicolumn{1}{c}{CE}&\multicolumn{1}{c}{EE}&\multicolumn{1}{c|}{Total}&\multicolumn{1}{c}{CE}&\multicolumn{1}{c}{EE}&\multicolumn{1}{c}{Total}\\ 
			\hline
			Numeric& Yes&0.000&0.000&0.000&0.000&0.000&0.000&0.000&0.000&0.000\\
			\hline
			Variational&Yes&0.015&0.011&-0.004&0.027&0.034&0.008&0.032&0.067&0.034\\
			Densities&No&0.014&0.014&0.000&0.027&0.034&0.008&0.012&0.065&0.049\\
			\hline
			Variational& Yes&-0.011&-0.008&0.002&-0.037&-0.033&0.001&-0.029&-0.029&-0.002\\
			Coulomb&No&0.002&0.004&0.002&-0.022&-0.019&0.001&-0.019&-0.019&-0.002\\
			\hline
			Variational& Yes&-0.015&-0.003&-0.008&0.044&0.034&-0.008&0.018&0.010&-0.007\\
			Polarization&No&0.00&0.011&-0.007&0.051&0.041&-0.007&0.024&0.019&-0.003
		\end{tabular}
	\end{ruledtabular}
	\label{Table2}
\end{table*}

\begin{table*}[t]
	\caption{The first-order XC energy error in Hartree due to fitting for atomic Mn
		and Zn using the three functionals.}
	\centering
	\begin{ruledtabular}
		\begin{tabular}{l c| d d d| d d d}
			&&\multicolumn{3}{c|}{Mn}&\multicolumn{3}{c}{Zn}\\
			\cline{3-8}
			Fitting
			Method&Constraint&\multicolumn{1}{c}{VWN}& \multicolumn{1}{c}{PBE}&\multicolumn{1}{c|}{BLYP}&\multicolumn{1}{c}{VWN}& \multicolumn{1}{c}{PBE}&\multicolumn{1}{c}{BLYP}\\ 
			\hline
			Variational&Yes&0.057&-0.042&-0.314&0.065&-0.398&-1.001\\
			Densities&No&0.051&0.071&0.020&0.065&0.039&-1.317\\
			\hline
			Variational& Yes&0.057&-0.042&-0.314&0.173&0.238&0.138\\
			Coulomb&No&0.013&-0.043&-0.462&0.152&0.218&0.115\\
			\hline
			Variational& Yes&0.054&0.118&0.098\\
			Polarization&No&0.046&0.105&0.061\\
		\end{tabular}
	\end{ruledtabular}
	\label{Table3}
\end{table*}

To begin to determine the properties of variational XC density fitting based on
Eq.~(\ref{Eq16}), atomic, all-electron calculations were performed for
diamagnetic Zn($^\mathrm{1}$S) and paramagnetic Mn($^\mathrm{6}$S), with five unpaired spins.   These atoms were chosen because both are spherically symmetric so that angular grids are not
needed in the numerical integration.  The parameter-free 80-point radial grid of
K\"oster, et al.\ \cite{Koester2004-2} was chosen, because no energy changed from that obtained using
a 70-point grid to the accuracy of the following tables.   Potentials and second
derivatives of the XC energy were obtained with the libxc library \cite{Marques2012}.  The
VWN LDA\cite{Vosko1980} and the PBE \cite{Perdew1996,Perdew1996Err} and BLYP \cite{Becke1988,Lee1988,Miehlich1989} GGA functionals were chosen.  For these two
atoms, variational density fitting was implemented in a completely numerical
fashion over the radial grid of 80 points, and involving the inverse of an 80 by
80 matrix for each functional and atom.  It should not be a surprise that the fitted and exact densities agree to
machine precision.  Thus variational density fitting works, but it is important
to begin to understand the effects of incomplete basis sets.

Atomic Zn calculations used the Turbomole triple-zeta plus polarization (tzp)
orbital basis \cite{Schafer1994},  which is contracted from a primitive17s/10p/6d orbital basis. 
The fitting basis set was created from the {\it s}-orbital-basis set with every exponent
doubled \cite{Dunlap1979,Dunlap1979-2},  which picks up the diagonal parts of the uncontracted orbital-product basis set. 
As this basis is not optimized for computing the CE, it would not be expected to
favor CE over CE plus XC fitting.

This particular orbital angular-momentum basis for Zn has two exponents from different contractions that are fairly close together, so that the smallest eigenvalue of the Coulomb metric matrix is of order 10$^\mathrm{-9}$.  Standard
algorithms give stable fits.  With a completely numerical treatment of the exact
XC energy the total PBE energy is -1779.12123\,Hartree.  One can separate the
fitting exponents by even tempering the set between the smallest and largest.  
The even-tempered fitting basis gives an energy that is 0.0023 H below that
using the orbital-derived fitting basis set.  Thus it is less accurate, because
the exact CE bounds the fitted CE from above.  Even tempering is not used
further.   An identical calculation can be performed using the smaller,
15s/9p/5d primitive, DGauss DZVP2 orbital basis set \cite{Andzelm1992}.  Its energy lies 0.3315\,H
higher, consistent with the variational principle.    These energy differences
set an appropriate energy scale for this work.   Any method with an error of
less than the middle ground of $\pm$0.03\,H would be expected to provide useful
transition-metal quantum chemistry and is taken as an arbitrary standard of
accuracy for this work.

The total electronic CE for atomic Zn is about 775\,H, and the XC energy is about
a tenth of that.  The kinetic plus nuclear attraction energy is about three
times the CE.  Table~\ref{Table1} gives total energy differences for Zn relative to the
calculation in which the CE is from an unconstrained Coulomb fit and the XC
energy is obtained numerically (numeric) using the exact density.   Compared to
that calculation, the first row gives the difference when the CE fit is
constrained.   In that case, the robust fitted CE and the total EE energy (EE)
must decrease by the Coulomb variational principle.  They do by 0.002\,H.  SCF
makes up for that difference, however, and the robust total DFT energy is
unchanged for all three functionals to the number of digits displayed in the
table.

In the second and third rows, the fully variational fitted density is used to
evaluate all energies.  The variational equations are solved by NR.  Thus third
derivatives of the robust XC energy with respect to each variational parameter
in the fitted density are required for these rows.   Third LDA derivatives are
provided by libxc.  Third GGA derivatives were obtained numerically from the
second derivatives that are provided.  Because the XC energy is negative, the NR
"metric" matrix at each SCF cycle is not positive definite, and Cholesky
decomposition cannot be used to find its inverse.   No problems with finding an
inverse were seen.

Our accuracy standard flags three of the four GGA total energies.   It says for
PBE an unconstrained fit is good, but a constrained fit is bad, but not very
bad.   For BLYP, both errors are significantly larger.   Fitting the BLYP
density is significantly harder than fitting the PBE density.   All fitting
errors are positive.   If they were negative, they could be attributed to the
Coulomb fit, but even the CE errors have the opposite sign for the two
functionals.  With full variational fitting, the tail, XC, is to a degree
wagging the dog, EC, for the GGA functionals, particularly BLYP.  No total
energy difference seen in Table~\ref{Table1} is within a factor of two of the total energy
difference using the two orbital basis sets.  Thus, fitting the density is never as bad as using a smaller orbital basis set.

The fourth row of Table~\ref{Table1} corresponds to the model-potential ParaGauss method,
which constrains the fitted density.   The fifth row corresponds to the ADFT
method, and the density is not constrained.   In both Coulomb-fitting methods
for all three functionals the CE and XC differences are essentially the same. 
Note that the robust XC fitted energy is used, i.e, the integral of the fitted
XC potential with the exact density is required, which modifies the ParaGauss and
ADFT energies.   This first-order energy is not computed in these
second-generation variational-fitting codes, but is fairly easy to add.  For non-spin-polarized
systems the methods are almost identical and quite accurate with the first-order
correction.

Atomic Mn calculations used the refined \cite{Calaminici2007}  double-zeta polarization (dzp) deMon2k
orbital basis (contracted from 15s/9p/5d).   Because of spin polarization, there are
roughly twice as many variational parameters in the fitted density.   Table~\ref{Table2}
displays all the Mn energy differences for the calculations that correspond to
those of Table~\ref{Table1}.  The last two rows were absent in the Zn Table.  In them, the two independent density variables are the nonzero spin
polarization and total density.   Because the spin-polarization of
the density only affects XC, it was fitted using a now negative-definite metric
matrix.  Of all
the Mn energies, only both PBE variational total energies are flagged by the
accuracy standard, but they are just barely over that standard.

The final table gives the robust correction to the fitted XC energies.   All
entries in Table~\ref{Table3} are greater in absolute value than the accuracy standard
except for the unconstrained, variational, Mn, BLYP and unconstrained,
CE-fitted, VWN density.   Four energy differences are even bigger than the
energy difference between the two orbital Zn basis sets.   Clearly, reliability
requires a robust fitted energy even for the relatively small XC energy of
transition metals.

\section{Conclusions}

The electron-electron energy of electronic structure methods is non-linear in the electronic density and/or orbital-pair charge distributions.  For HF, this energy is quadratic in orbital-pair charge distributions, for the Slater XC energy \cite{Slater1951}, it goes as the density to the 4/3 power, and for second-order HF it goes like orbital-pair charge distributions to the fourth power.  The general GGA energy has a more complicated dependence on density, which again, is non-linear in the density and its gradient.  

The variation of the energy with respect to the set of all charge distributions that contribute to the EE energy generates a set of potentials.  These can be combined to collectively generate the total EE potential.  This potential is operationally similar to, and maybe even should be taken as, the KS potential of modern DFT.    Generating the potential from approximate charge distributions that are written as sums over single-center Gaussian basis functions simplifies computation.

As with KS DFT, the expectation value of this potential is not the energy but requires a DC correction.  Having both the original and fitted charge distributions makes any expression for energy ambiguous because it is not clear which representation of the charge distributions should be used in which of the various constituent pieces.   Requiring the variation of the energy with respect to the exact charge distributions to produce the fitted set of potentials removes this ambiguity while leaving computation simple.  Variational fitting makes both the potential and the DC correction functions of the fitted charge distributions.

This fitted energy is correct to first-order for differences between each charge distribution and its fit.
Applying the CoV to make this robust energy independently stationary with respect to the orbitals and the fitted density results in two coupled sets of equations that determines both sets of quantities.  Variational fitting couples all fits because each fit affects the EE potential.

We have, for the first time, defined a robust expression for the entire EE interaction energy and determined the corresponding coupled set of potential fitting equations, Eq.~(\ref{Eq16}).  We have described variational fitting for HF through second order and DFT.  For DFT this simplifies previous results for variational fitting of the Coulomb potential and Slater XC potential, which involves separate fits of the density and Slater exchange potential \cite{Dunlap2011} at the price of numerical integration. The formalism extends up the entire DFT ladder of approximations.  These robust energies could be used to generate optimized fitting basis sets.  That would be an improvement over existing empirical fitting basis sets, because the super linear nature of the EE interaction energy couples all fitted quantities, the orbitals and each density variable of each rung of Perdew's ladder.

The largest errors seen in our preliminary study of fully variational density fitting in DFT are the first-order errors needed to make the fitted energy robust.  Thus fully variational density fitting seems practical.  The first-order error is not zero even when the density fit is not constrained.  In all cases, the first-order XC error is shown to be significant in this work.  Nevertheless, there is cancellation between the CE and XC errors, and one can neglect the XC error if the first-order CE error is also neglected.   The variational principle is balancing CE and XC errors, even though CE is an order of magnitude larger.  In all cases the EE energy is approximately equal to an expression that contains no orbitals,
\begin{equation}
CE+E_{XC}=\langle\bar\rho|\bar\rho\rangle+\int\varepsilon_{XC}\left[\bar\rho_\uparrow,\bar\rho_\downarrow\right]d\mathbf{r}
\label{Eq22}
\end{equation}
Orbitals only appear in the fitting Eq.~(\ref{Eq16}) that is needed to
determine the fitted potentials.  This robust EE energy is numerically stationary, as required by the HK and KS theorems.   The calculations of the Tables
were actually performed taking Eq.~(\ref{Eq16}) as primary.  Therefore, only the
unconstrained variational density calculation precisely satisfies
this equation.

Six different robust approaches to KS DFT using fitting basis sets were
studied using three different functionals.   Variational density fitting works extremely well
with the LDA.  In all cases, the first-order XC error, included in
Eq.~(\ref{Eq14}) must be used.  Direct atomic {\it s} fitting basis-set optimization
is probably necessary for the GGA given an orbital basis.   Better yet, variational fitting allows simultaneous orbital and fitting basis set optimization, which would generalize the Pople orbital basis sets to the entire set of basis functions used in a calculation.  

It is possible to fit 
the entire EE energy in a fully variational fashion for all of DFT.   Due to the fact that the GGA is quite complex yet can be variationally fitted, it is likely that all electronic-structure models can be transformed.  The solution of the coupled, fitted, and variational potentials of Eq.~(\ref{Eq16}) can be inserted directly into the Fock matrix for orbital generation.  Then the exact ladder densities are put into Eq.~(\ref{Eq16}) to complete the SCF cycle.

\begin{acknowledgments}
We thank Marcin Dulak for help in linking libxc on a Mac.  This work is supported by the Office of Naval Research, directly and through the Naval Research Laboratory.  M.C.P. gratefully acknowledges an NRC/NRL Postdoctoral Research Associateship.
\end{acknowledgments}

\bibliography{citations}{}
\end{document}